\documentclass[pra,superscriptaddress,reprint]{revtex4-1}
\usepackage{graphicx}
\usepackage{enumitem}
\usepackage{amsmath}
\usepackage{bbm}
\usepackage{color}
\newcommand{\ket}[1]{|{#1}\rangle}

\newcommand{\cm}[1]{\ignorespaces}

\usepackage{amsfonts}

\usepackage[usenames,dvipsnames]{xcolor}
\usepackage[colorlinks=true,citecolor=Cerulean,linkcolor=RubineRed,urlcolor=Cerulean]{hyperref}

\begin{document}

\title{Linked and knotted synthetic magnetic fields}

\author{Callum W. Duncan}
\affiliation{SUPA, Institute of Photonics and Quantum Sciences,
Heriot-Watt University, Edinburgh EH14 4AS, United Kingdom}
\author{Calum Ross}
\affiliation{Maxwell Institute for Mathematical Sciences and Department of Mathematics, Heriot-Watt University, Edinburgh EH14 4AS, United Kingdom}
\author{Niclas Westerberg}
\affiliation{SUPA, Institute of Photonics and Quantum Sciences,
Heriot-Watt University, Edinburgh EH14 4AS, United Kingdom}
\author{Manuel Valiente}
\affiliation{Institute for Advanced Study, Tsinghua University, Beijing 100084, China}
\affiliation{SUPA, Institute of Photonics and Quantum Sciences,
Heriot-Watt University, Edinburgh EH14 4AS, United Kingdom}
\author{ \\ Bernd J. Schroers}
\affiliation{Maxwell Institute for Mathematical Sciences and Department of Mathematics, Heriot-Watt University, Edinburgh EH14 4AS, United Kingdom}
\author{Patrik \"{O}hberg}
\affiliation{SUPA, Institute of Photonics and Quantum Sciences,
Heriot-Watt University, Edinburgh EH14 4AS, United Kingdom}

\begin{abstract}
\noindent We show that the realisation of synthetic magnetic fields via light-matter coupling in the $\Lambda$-scheme implements a natural geometrical construction of magnetic fields, namely  as the pullback of the area element of the sphere to Euclidean space via certain maps. For suitable maps, this construction generates linked and knotted  magnetic fields,  and the synthetic realisation amounts to the identification of the map with the  ratio of two Rabi frequencies which represent the coupling of the internal energy levels of an ultracold atom. We consider examples of maps which can be physically realised in terms of Rabi frequencies and  which lead to linked and knotted synthetic magnetic fields acting on the neutral atomic gas. We also show that the ground state of the Bose-Einstein condensate may inherit topological properties of the synthetic gauge field, with linked and knotted vortex lines appearing in some cases.
\end{abstract}

\maketitle

\section{Introduction}

Lord Kelvin's conjecture 150 years ago that atoms are made of knotted vortex structures \cite{thomson1867} anticipated today's 
study, both theoretical and experimental, of  topological structures in nature. Non-trivial topological structures have been studied in
classical fluids \cite{moffatt1969,moffatt1992,kleckner2013,enciso2013,Scheeler2014}, plasma physics \cite{Chandrasekhar1958,Berger1999,Thompson2014}, nuclear physics \cite{Battye1998,Sutcliffe2007}, condensed matter physics  \cite{Sutcliffe2017}, DNA \cite{taylor2000,suma2017}, soft matter \cite{tkalec2011} and light \cite{irvine2008,irvine2010,Kedia2013,kedia2017}. There has also been interest in the physics of topological magnetic field lines, with research focusing on their construction \cite{kedia2016} and stability \cite{kleckner2013,Scheeler2014,kleckner2016}. Understanding the behaviour of matter in non-trivial topological magnetic fields is also important in the study of plasma physics and the determination of stable confining magnetic field configurations in thermonuclear reactors \cite{moffatt2014,moffatt2015}. 

Ultracold atoms allow the realisation of synthetic gauge fields in such a way that neutral atoms mimic the dynamics of charged particles in a magnetic field \cite{Spielman2009,Lin2009,Lin2011,Zhu2006,Dalibard2011,goldman2014}. One method of creating a synthetic gauge field is to exploit atom-light couplings by driving internal transitions of the atoms to realize static Abelian gauge fields which are tunable via the applied laser \cite{juzeliunas2004,juzeliunas2005,juzeliunas2006}.  There has also been significant interest in creating knotted structures in quantum gases \cite{Liang2009,Liu2013,Liu2014,Proment2014,Bidasyuk2015}, with the first knots in quantum matter having been realised in spinor BECs \cite{Kawaguchi2008,hall2016}. Further experiments have investigated the formation of a  Shankar skyrmion in the spinor BECs with knotted spin structure \cite{Shankar1977,Lee2018}. The imprinting of linked and knotted vortex structures has also been proposed using driving schemes of the internal energy levels \cite{Ruostekoski2005,Maucher2016}.

The formation of knotted vortex lines (or knotted solitons) has been extensively investigated in superfluids and superconductors \cite{Babaev2002,Babaev2002b,Babaev2004}, including the conditions for their stability in multicomponent superconductors \cite{Rybakov2018}. In addition, there have been proposals for fault-tolerant \cite{Shor1995} topologically protected quantum computations  \cite{Kitaev2003,Nayak2008} using vortices in superconductors \cite{Read2000,Sarma2006} and spin interactions in optical lattices \cite{Jaksch1999,Duan2003}. The knotted vortices considered in superconductors are of a different nature from those we consider here. However, as we will explain in Sec.~\ref{sec:Conclusions}, that knotted or linked magnetic fields generically open up new avenues for quantum computing.

In this paper, we point out and exploit a remarkably direct link between the realisation of synthetic magnetic fields in ultracold atoms and a mathematical construction of knotted and linked magnetic fields,  due to Ra\~{n}ada \cite{ranada1989,Ranada1990},  out of a map from Euclidean 3-space to the 2-sphere. In a nutshell, we show that this map can be realised  as the ratio of two complex Rabi frequencies describing the atom-light coupling in a three-level atomic $\Lambda$-scheme. 

The mathematically most natural choice of the Ra\~{n}ada map for a given 
link or knot is challenging to implement directly in an experiment, 
but our results suggest that one  can  implement an approximation to this map which, crucially, preserves the topology of the knot or link.  
We propose a general method for constructing this approximation, and  illustrate it with three examples of maps, called  $f_H,f_L$ and $f_T$,  whose associated magnetic field lines are, respectively, Hopf circles, linked rings and the  trefoil knot.

Finally, we find that some of the topological structure of the synthetic gauge field lines  is  inherited by the ground state of the dark-state wavefuntion in the $\Lambda$-scheme. The details of this depend on the potential and magnetic field  which appear in the effective Hamiltonian. For instance, if the scalar potential is peaked along  a knot or link, the wavefunction reflects this through a vortex structure along that knot or link. This happens for the trefoil knot and  linked rings, but not for the Hopf circles, where the  scalar potential is spherically symmetric. A similar interplay between linked magnetic field lines and vortex lines in a spinorial wavefunction was recently studied in Ref. \cite{Ross2018}.

\section{Ra\~{n}ada's knotted light} 

In the 1980s, Ra\~{n}ada proposed a systematic mathematical construction of linked or knotted electric and magnetic fields \cite{ranada1989,Ranada1990}. This construction has a beautiful geometrical interpretation in terms of the  geometry of the 2-sphere which we review briefly in Appendix~\ref{app:Ranada}.  It leads to an explicit formula for a magnetic field in terms of a map $f: \mathbb{R}^3\rightarrow S^2$. Identifying  the   2-sphere with $\mathbb{C} \cup\{ \infty\}$ via the stereographic projection, the magnetic field is
\begin{equation}
\mathbf{B} = \frac{1}{2 \pi i} \frac{\nabla f^* \times \nabla f}{\left(1 
+\lvert f \rvert^2\right)^2}.\label{eq:RanadaMag}
\end{equation}
This field has vanishing divergence, and therefore satisfies the static Maxwell equations, generally with a  non-trivial current. Moreover, one checks that field lines are determined by the (complex) condition $f=$ constant. In this way one can therefore construct topologically interesting magnetic fields by drawing on the extensive mathematical literature studying links and knots as level curves of complex functions. The formulation of the magnetic field in Eq.~\eqref{eq:RanadaMag}, along with the definition of the maps, has been utilised to study the properties of topologically non-trivial vector fields \cite{irvine2008,Besieris2009,Kedia2013,Thompson2014,kedia2016,kedia2017}.

The maps $\mathbb{R}^3 \rightarrow S^2$ considered by Ra\~{n}ada and in this paper
go via $S^3$, i.e. they are compositions 
\begin{equation}
f: \mathbb{R}^3 \rightarrow S^3 \rightarrow S^3 \rightarrow S^2,
\label{compmap}
\end{equation} 
  where the first step is the inverse stereographic projection in three dimensions, and the last step is projection of the Hopf fibration. The details of the map are encoded in the intermediate step $S^3 \rightarrow S^3$. Using complex coordinates $u,v \in \mathbb{C}$ satisfying  $\lvert u \rvert^2 + \lvert v \rvert^2 =\mathit{l}^2 $  to parametrise the 3-sphere of radius $\mathit{l}$, the inverse stereographic projection maps $(x,y,z) \in \mathbb{R}^3$ to
\begin{equation}
u = \frac{2 \mathit{l}^2 \left(x+iy\right)}{\mathit{l}^2+r^2}, \: \: v = \frac{2 \mathit{l}^2 z+i \mathit{l} \left(r^2-\mathit{l}^2\right)}{\mathit{l}^2+r^2},\label{eq:uandv}
\end{equation}
with $r^2 = x^2 + y^2 + z^2$
and $\mathit{l}$ setting the unit of length. The map \eqref{compmap} then takes the form 
\begin{equation}
f\left(x,y,z\right) = \frac{g\left(u\left(x,y,z\right),v\left(x,y,z\right)\right)}{h\left(u\left(x,y,z\right),v\left(x,y,z\right)\right)},
\label{fexplicit}
\end{equation}
where $g$ and $h$ are complex functions of  $u$, $v$ which  must not  vanish simultaneously.

In this paper  we focus on three examples; 
the standard  Hopf map $f_H = u/v$ (defining Hopf circles),  the quadratic Hopf map  $f_L = u^2/(u^2-v^2)$ (defining linked rings) and the map $f_T = u^3/(u^3+v^2)$ which defines the trefoil knot. The  first two define  links whose topology is independent of the chosen (complex) level; the level curves of $f_H$ are circles or the $z$-axis (an ``infinite circle"), and any two  circles link once.  The level curves of $f_L$ are linked rings or the $z$-axis;  different level curves link each other four times. 
The third map defines a trefoil  knot for level  $\infty$ or sufficiently large.

\begin{figure}[t]
\begin{center}
\includegraphics[width=0.35\textwidth]{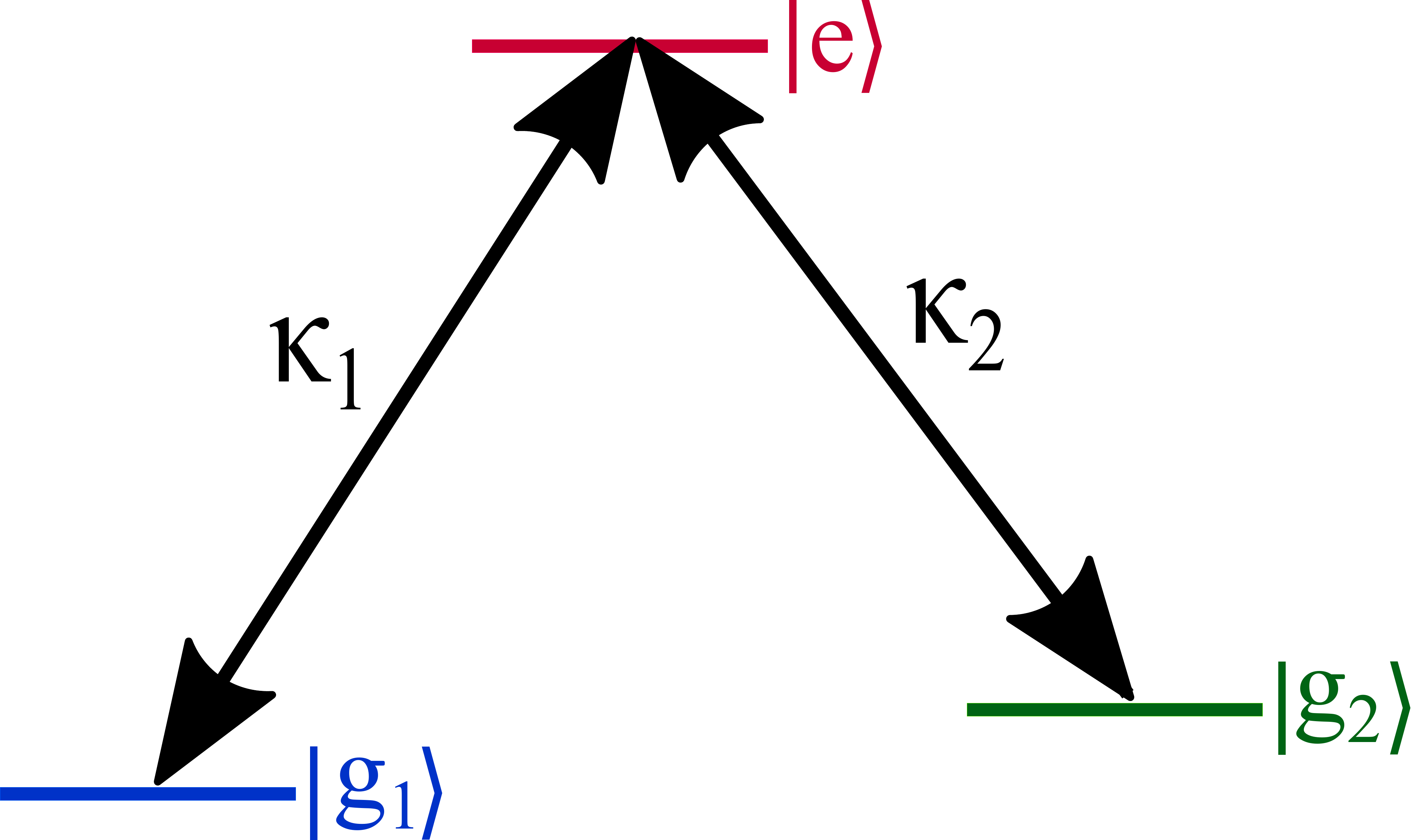}
\end{center}
\caption{Illustration of the $\Lambda$-scheme, with internal atomic energy levels $\ket{e}$, $\ket{g_1}$ and $\ket{g_2}$ coupled by lasers $\kappa_1$ and $\kappa_2$.}
\label{fig:LambdaScheme}
\end{figure}

\section{The $\Lambda$-scheme} 

Synthetic gauge potentials for ultracold atoms can be realised in many ways \cite{Dalibard2011,goldman2014}. We  consider an ensemble of atoms with three internal energy levels
where two ground states $|g_1\rangle$ and $|g_2\rangle$ are coupled by two laser beams to a third excited state $\ket{e}$. This configuration of energy levels is called a $\Lambda$-scheme, and is illustrated in Fig. \ref{fig:LambdaScheme}. The strength of the atom-light coupling is characterised through space-dependent, complex Rabi frequencies $\kappa_1,\kappa_2$.  We assume the lasers are resonant with the transitions, and with zero two-photon detuning, resulting in the  atom-light coupling Hamiltonian
\begin{equation}
H_{\mathrm{int}}=   \left(\begin{matrix} 
      0 & 0 & \kappa_1 \\
      0 & 0 & \kappa_2\\
      \kappa_1^* & \kappa_2^* & 0
   \end{matrix}\right).\label{hint}
\end{equation}
A general state of the light-matter coupled system can then be written as $|\Psi\rangle=\sum_{i=D,+,-}\psi_i(x)|i\rangle$  where $|D\rangle,|+\rangle,|-\rangle$ depend parametrically on space  and are the three eigenstates of 
$H_{\mathrm{int}}$. The eigenstate for eigenvalue zero is the dark state
 \begin{equation}
 \ket{D} = \frac{1}{\sqrt{|\kappa_1|^2+|\kappa_2|^2}}\begin{pmatrix}\phantom{-} \kappa_2^* \\  -\kappa_1^*\\ \phantom{-}0\end{pmatrix}.
 \end{equation}
It has no contribution from the excited state and is therefore also robust against detrimental spontaneous decay.

\begin{figure*}[t]
\begin{center}
\includegraphics[width=0.8\textwidth]{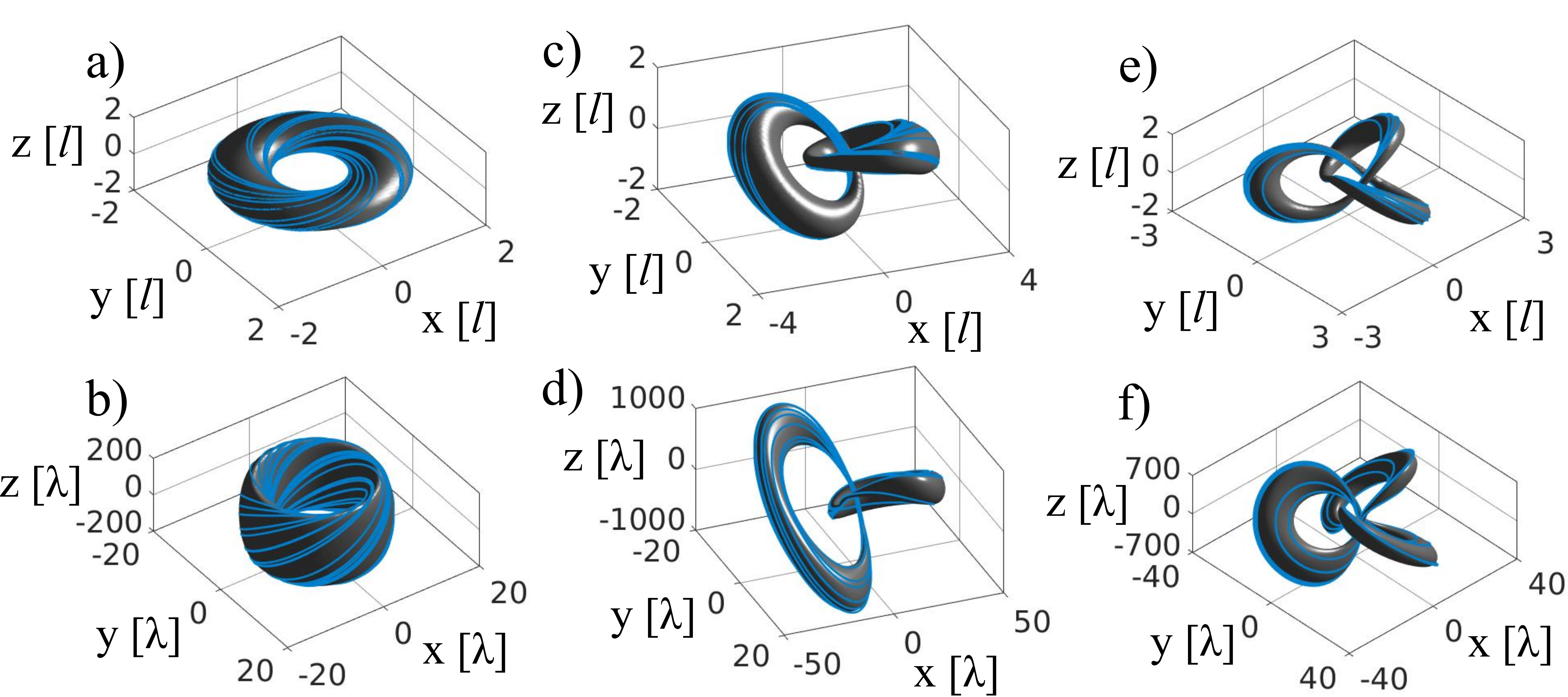}
\end{center}
\caption{ Exact and approximated magnetic field lines,  realised  as level curves of the complex field $f$ and its Laguerre-Gauss approximation $\zeta$ .  We show level surfaces   of  $|f|$ and $|\zeta|$,   and, on each level surface, we show magnetic field lines in light blue. a) Exact Hopf circles ($f_H$). b) Realised Hopf circles ($\zeta_H$). c) Exact linked rings ($f_L$). d) Realised linked rings ($\zeta_L$). e) Exact trefoil knot ($f_T$). f) Realised trefoil knot ($\zeta_T$). The unit of length for the exact magnetic fields is $l$ and for the realised fields it is the laser wavelength $\lambda$ with $\alpha = 100$.}
\label{fig:FieldLines}
\end{figure*}

If we include the kinetic term  and a confining potential $V$ in the full Hamiltonian $H=\frac{\mathbf{p}^2}{2m}+H_{\mathrm{int}}+V$ and, using the adiabatic approximation,  project the corresponding Schr\"odinger equation $i\hbar\partial_t|\Psi\rangle=H|\Psi\rangle$ onto the dark state  while neglecting the coupling to the other dressed states, then $\psi_D$ is governed by the equation of motion 
\begin{equation}
i\hbar\frac{\partial}{\partial t} \psi_{D} = \left[ \frac{\left(\mathbf{p}- \mathcal{A}\right)^2}{2m} + \Phi + V \right] \psi_{D}.
\label{eq:eom}
\end{equation}
 The vector potential $\mathcal{A}$, the corresponding magnetic field $\mathcal{B}$ and geometric potential $\Phi$ are fully determined by the Rabi coefficients $\kappa_1, \kappa_2$, with the magnetic field and scalar potential conveniently expressed in terms of 
 $\zeta = \kappa_1/\kappa_2$.
Explicitly we have
\begin{equation}
\mathcal{A} =  \frac{ i \hbar( \kappa_1 \nabla \kappa^*_1 + \kappa_2 \nabla \kappa^*_2 -  \kappa^*_1 \nabla \kappa_1  -  \kappa^*_2 \nabla \kappa_2)}
{2(|\kappa_1|^2 + |\kappa_2|^2)}, 
 \label{eq:LambdaA}
 \end{equation}
 \begin{equation}
 \mathcal{B} =  i \hbar \frac{\nabla \zeta \times \nabla \zeta^*}{\left(1+\lvert \zeta \rvert^2\right)^2},\label{eq:LambdaB}
  \end{equation}
  \begin{equation}
\Phi =   \frac{\hbar^2}{2 m} \frac{\nabla \zeta^*  \cdot  \nabla \zeta }{\left(1+\lvert \zeta \rvert^2\right)^2}.
\end{equation}
Note, that expressing $\mathcal{A}$ in terms of $\zeta$ would lead to a singular gauge. 

For the synthetic magnetic field and its corresponding gauge potential to be experimentally viable, the life-time of the dark state needs to be long enough. 
For example, spontaneous emissions from the excited state would change the life time of the dark state, but this is mitigated if the Rabi frequency is large enough to ensure the adiabatic approximation is valid. In addition, strong collisional interactions should be avoided, as any atom-atom interactions will be detrimental to the stability of the dark state. This can be addressed by ensuring the atoms are in the dilute limit or the scattering length is tuned to be small. We also require that any Zeeman coupling terms between the two ground states of the $\Lambda$-scheme are sufficiently small to allow them to be neglected.

If we identify $\zeta \equiv f$ then the magnetic fields of Ra\~nada, Eq.~\eqref{eq:RanadaMag}, and the $\Lambda$-scheme, Eq.~\eqref{eq:LambdaB}, are equivalent (in fact, equal with $\hbar=1/2\pi$). Therefore, to realize the topological magnetic field of a particular $f$ we are required to drive the atomic transitions by the Rabi frequencies 
 such that their ratio, $\zeta$, forms the mapping $f$. The Rabi frequencies $\kappa_1$ and $\kappa_2$ can be chosen independently, giving  a considerable amount of flexibility and allowing us, in principle, to realise any link or knot which is the level curve of a function $f: \mathbb{R} \rightarrow S^2$. 
However, we can not set the Rabi frequencies to be any arbitrary function of space and phase, as they are realised by laser beams which need to fulfil Maxwell's equations. This restriction on the allowed forms of the Rabi frequencies is at the heart of our discussion in the next section.

\section{Realisation of topological fields.} 

Our approach is inspired by  Refs.~\cite{dennis2010,Romero2011,Padgett2011}, where linked and knotted optical vortex lines were realised in laser beams as  a superposition of Laguerre-Gaussian (LG) modes. These superpositions of LG modes are usually obtained by the use of Spatial Light Modulators (SLMs) \cite{McGloin2003,Bergamini2004,boyer2006,olson2007,gaunt2013,nogrette2014,Bijnen2015}. LG beams are characterised by their azimuthal, $n$, and radial, $p$, indices, and we will denote a single LG mode as $\mathcal{L}_{pn}$, with the full definition of a LG mode discussed in Appendix~\ref{app:LG}. Our method of constructing the topological synthetic magnetic fields consists of the following steps:
\begin{enumerate}
\item Starting from a  map $f$  of the form \eqref{fexplicit}, restrict it  to the $z=0$ plane, where $z$ is the direction of propagation for the lasers, and note that the result is a ratio of polynomials $p$ and $q$ in $x$ and $y$.
\item Expand the polynomials $p$ and $q$  in terms of  LG modes restricted to the $z=0$ plane  and without the common Gaussian factor.
\item Replace $p$ and $q$ in $f$ by the expansions in the LG modes, including their $z$-dependence (and note that the common Gaussian factor cancels).
\item Check numerically if the level curves of the resulting function $\zeta$ have  the same  topology as  the  level curves of $f$.
\item 
If they do, realise the level curves   as synthetic magnetic field lines  via $\zeta= \kappa_1/\kappa_2$, where $\kappa_1  $ and $\kappa_2$ are the LG modes approximating $g$ and $h$. 
\end{enumerate}
All three examples considered in this work pass the check in step 4, but we are not aware of a  mathematical proof  that this should be true 
generally.

\subsection{Form of the topological magnetic fields} 

Expanding in terms of LG modes we find the following approximations for the Hopf map $ f_H$, the quadratic Hopf map $f_L$ and the map $f_T$ for the trefoil knot:
\begin{equation}
\zeta_H =  \frac{2 \alpha \mathcal{L}_{0 -1}}{i\left[\left(\frac{\alpha^2}{2}-1\right)\mathcal{L}_{00} - \frac{\alpha^2}{2} \mathcal{L}_{10}\right] },
\label{eq:ZetaHopf}
\end{equation}
\begin{equation}
\zeta_L =  \frac{(2 \alpha)^2 \mathcal{L}_{0 -2}}{(2 \alpha)^2 \mathcal{L}_{0 -2} + c_0 \mathcal{L}_{00} + c_1 \mathcal{L}_{10} + c_2 \mathcal{L}_{20}},
\end{equation}
\begin{equation}
\zeta_T =  \frac{(2 \alpha)^3 \mathcal{L}_{0 -3}}{(2 \alpha)^3 \mathcal{L}_{0 -3} + c^\prime_0 \mathcal{L}_{00} + c^\prime_1 \mathcal{L}_{1 0} + c^\prime_2 \mathcal{L}_{2 0} + c^\prime_3 \mathcal{L}_{3 0}},
\label{eq:ZetaTrefoil}
\end{equation}
where we have defined $\alpha = \omega_0 / l$ with $\omega_0$ the beam waist of the laser. Definitions of the coefficients $c_i$ and $c^\prime_i$, which are polynomials in $\alpha$, can be found in Appendix~\ref{app:LG}.

A comparison of the exact and realised magnetic fields for all three cases considered is shown in Fig.~\ref{fig:FieldLines}.  For all realised fields we have chosen a beam width of $\alpha=100$ 
and work in units of the wavelength of the laser $\lambda$.  The realised fields are found to be stretched out in the $z$-direction compared to the exact fields. For all three examples considered the topological nature of the realised magnetic field lines is clear, as the level set of each have similar forms to that of the exact fields. 

\subsection{Beam shaping realisation} 

We note here that the specific combinations of LG modes discussed in the previous subsection is entirely a beam shaping excercise. That is, the superposition of the LG modes in the denominators of Eqs.~(\ref{eq:ZetaHopf}-\ref{eq:ZetaTrefoil}) results in a single beam with the required intensity and phase profiles. This single beam is then used as $\kappa_2$ in the driving of the $\Lambda$-scheme.

The atomic transitions accessed in a $\Lambda$-scheme are typically in the optical regime and thus the diffraction limit of $0.2~\mu m -~0.4~\mu m$  sets the 
length scale limit. Furthermore, the resolution of current beam shaping technology imposes  limits on the spatial resolution of the resulting gauge field and on the field strength.   The atomic cloud size ($\sim 100 \mu  m$ \cite{Jaouadi2010}) is typically smaller than the usual beam waists considered (e.g. $\sim 1mm$ \cite{Leach2005}). Nonetheless, we do not foresee that the gauge field configurations discussed here will fall outside what is currently experimentally achievable, as it is not unusual to focus optical beams down to beam waists of $50-200 \mu m$ in other settings \cite{Roger2016,Caspani2016}.

\begin{figure}[t]
\begin{center}
\includegraphics[width=0.49\textwidth]{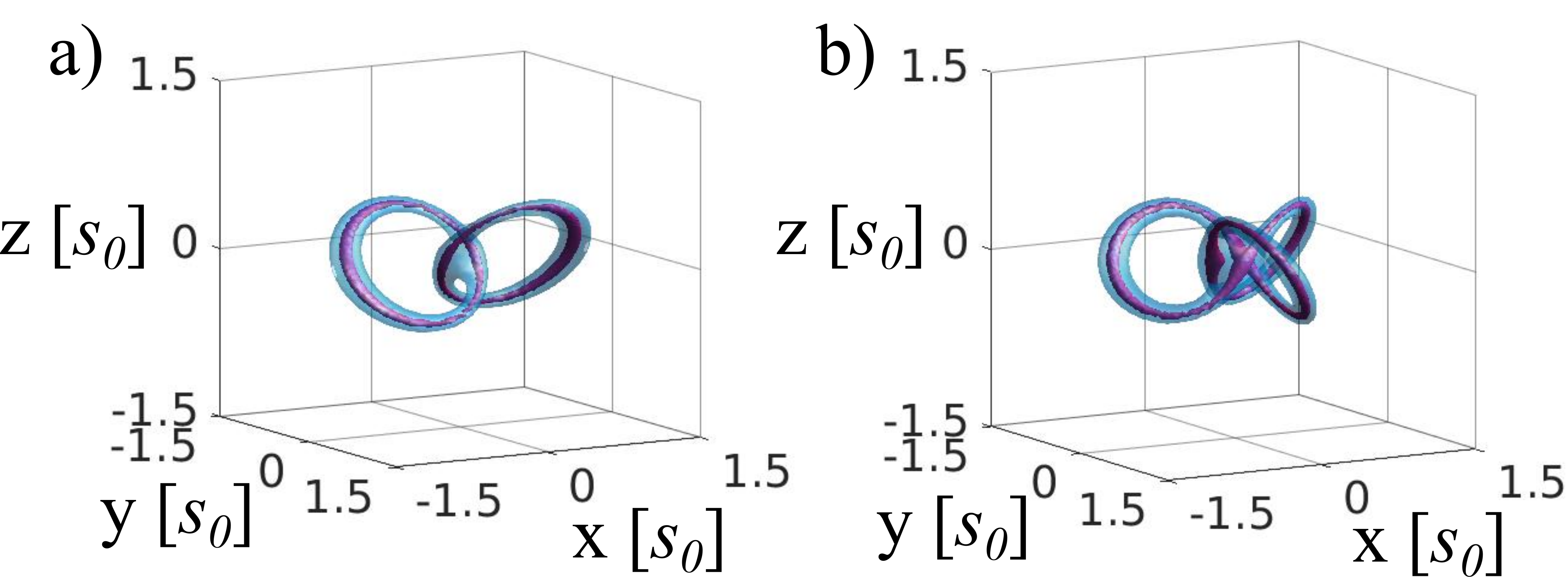}
\end{center}
\caption{Vortex structure of the ground states of the Hamiltonian in \eqref{eq:eom} with a vector potential constructed from a)  $f_L$ and  b) $f_T$.
Shown are level sets of the probability density ($|\psi|^2 = \mathrm{const.}$) for a small constant, which visualises the vortex core structure. The vortices form linked rings  in a) and a trefoil knot in  b), thus replicating the form of the magnetic field lines in both cases.}
\label{fig:GS}
\end{figure}

\subsection{Ground state of the quantum gas} 

In order to illustrate the effect the knotted synthetic magnetic fields can have on atoms, we envisage a non-interacting gas of atoms forming a three-dimensional Bose-Einstein condensate which is trapped by a harmonic external potential $V = m \omega^2 r^2/2$, where $\omega$ is the trap frequency. We are interested in the properties of the ground state of such a condensate which is interacting with a linked or knotted magnetic field and a geometric potential via Eq. \eqref{eq:eom}. We solve for the ground state $\psi = \psi_{\mathrm{D}}$ using imaginary time propagation \cite{Chiofalo2000,Roy2001,Bader2013} on a $201^3$ numerical grid and for the three exact gauge fields defined via the maps $f_{\mathrm{H}}$, $f_{\mathrm{L}}$ and $f_{\mathrm{T}}$.  We choose our unit of length to be $s_0 = \sqrt{\hbar/m\omega}$ and take $l=1$. It is not immediately obvious what the properties of the ground states of this system should be. The ground state is dependent on the interplay between the strength and shape of the topologically non-trivial magnetic field, the corresponding scalar potential, and the trapping potential. The ground states discussed here reflect the choice of considering a cloud of cold atoms confined in a harmonic potential.

\begin{figure}[t]
\begin{center}
\includegraphics[width=0.49\textwidth]{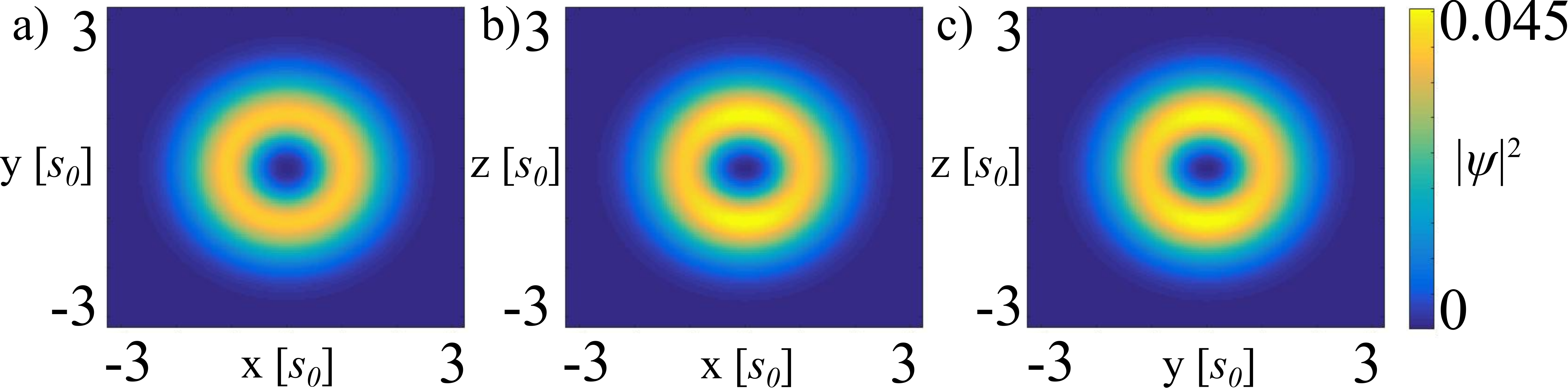}
\end{center}
\caption{Ground state of the Hamiltonian in \eqref{eq:eom}  with a vector potential constructed from $f_H$.
  The probability density is shown  in the a) $xy$-plane ($z=0$), b) $xz$-plane ($y=0$) and c) $yz$-plane ($x=0$). The ground state is real valued and forms a shell structure which is close to spherical but  slightly elongated in the $z$-direction. Note that the geometric potential $\Phi$ is spherically symmetric in this case, but the magnetic field is not}
\label{fig:Cuts}
\end{figure}

We observe the presence of vortex structures in the ground states for  the linked rings  and  the trefoil knot, which are shown in Fig.~\ref{fig:GS} by the level sets of $|\psi|^2$ and in the supplemental movies \cite{SuppMov}.  However, there is no vortex structure in  the ground state for  the Hopf circles
shown in Fig.~\ref{fig:Cuts}, for this choice of parameters. The vortex structures in the other ground states are determined by the maxima of the scalar potential  $\Phi$,  whose level sets for near-maximal values are very similar to the level sets for small probability density shown in Fig.~\ref{fig:GS}.  We are not aware of a simple  mathematical  reason for the relation between the level sets of $\Phi$ and the magnetic field lines which we observe for the linked rings and the trefoil knot. 

The detection of such topological structures in the gas requires a tomographic approach where 
3D vortex core structure are imaged using a non-destructive measurement of the density \cite{Higbie2005}. Alternatively, the presence of non-trivial gauge fields can 
be indirectly detected by measuring the shape oscillations of the gas \cite{Murray2007}.

\section{Conclusions} 
\label{sec:Conclusions}

We have shown that certain magnetic fields which are the  pullback of the normal area element of the 2-sphere to Euclidean 3-space can be realised as a synthetic  magnetic field  in the resonant $\Lambda$-scheme. 
Based on this observation, we propose a five-step method of realising general synthetic topological magnetic fields  using a superposition of LG modes.
We have derived the required LG superpositions for three examples -- the Hopf circles, the linked rings and the trefoil knot -- and shown their topological nature. In some cases, the topological form of these magnetic fields can 
be transferred to the ground states of the ultracold gas in the form of linked and knotted vortex cores. The general method presented in this work is not limited to the three examples considered,  and we expect 
 more links and knots defined by a  map $f$ to be realisable. 

The 3D-nature of the generated  states  and the versatility of our method opens a possible avenue for a physical realisation of the motion group of links or knots \cite{goldsmith1981,aneziris1991}. This group is a generalisation of the braid group of a surface, and includes elements which describe truly three-dimensional motions, for example  motions where  one  circle is pulled  through another. If such motions proceed through   configurations which are the level sets of a complex-valued function they  can,  in  principle, be realised in our scheme. The  quantum states of the condensate would potentially  pick  up exotic and non-abelian phases in such motions, reflecting the intricate (and little studied) representation theory of the motion group. It would clearly be interesting to investigate this possibility and to study its potential use in fault-tolerant topologically protected quantum computing. 

\begin{acknowledgments}
The authors thank Calum Maitland for helpful discussions. C.W.D. and N.W. acknowledge support from EPSRC CM-CDT Grant No. EP/L015110/1. C.R. acknowledges an EPSRC-funded PhD studentship. P.\"O. and M.V. acknowledge support from EPSRC EP/M024636/1.
\end{acknowledgments}


\appendix

\section{Differential geometry behind Ra\~{n}ada's construction}
\label{app:Ranada}

Ra\~{n}ada's construction of a magnetic field in terms of a map $f: \mathbb{R}^3\rightarrow S^2$ is  most easily stated in the language of differential forms and pull-backs.  This clarifies the coordinate-independent nature of the formula for the magnetic field, and  provides a basis for generalisations. We give a succinct summary here, referring the reader to textbooks like \cite{fecko2006} for the differential-geometric background.

A fundamental role is played by the  2-form representing the area element $\Omega$ of the 2-sphere.  Parametrising the 2-sphere  via stereographic projection  in terms of a complex coordinate $\mathcal{Z} \in \mathbb{C} \cup\{ \infty\}$, this 2-form is
\begin{align}
\Omega =  \frac{1}{2 \pi i} \frac{d \mathcal{Z}^* \wedge d 
\mathcal{Z}}{\left(1 + \lvert \mathcal{Z} \rvert^2 \right)^2}.
\end{align}
It  is manifestly closed, i.e. satisfies $d\Omega=0$, and normalised to unit area. Given a map $f: \mathbb{R}^3\rightarrow S^2$, Ra\~{n}ada's magnetic field is the pull-back
\begin{align}
f^*\Omega  =  \frac{1}{2 \pi i} \frac{d f^* \wedge d f}{\left(1 + \lvert 
f \rvert^2 \right)^2}
\end{align}
of $\Omega$ with $f$. This pull-back is a 2-form on $\mathbb{R}^3$ and automatically closed: it satisfies $d(f^*\Omega)= 0$ because pull-back commutes with the exterior derivative. The magnetic field  $\mathbf{B} $ given by Eq. (1)  in the main text is the vector field associated to $f^*\Omega$ using the metric and volume element of Euclidean space. The closure of $f^*\Omega$ is then equivalent to $\mathbf{B} $ having vanishing divergence.

\begin{widetext}

	\section{Laguerre-Gaussian expansion technique}
\label{app:LG}

We provide the details for  the expansions of the three example maps 
\begin{align}
f_H=  \frac{u}{v}, \quad f_L= \frac{u^2}{u^2-v^2}, \quad f_T = \frac{u^3}{u^3+v^2},
\end{align} 
considered  in the main text  in terms of the complete  set of Laguerre-Gaussian (LG) beams, following the five-step method also proposed in the main text. 
The  LG modes are
\begin{align}
\mathcal{L}_{pn}\left(\rho,\phi,z\right) = \frac{C}{\sqrt{1+\frac{z^2}{z_R^2}}}\left(\frac{\rho \sqrt{2}}{w\left(z\right)}\right)^{|n|} L^{|n|}_p \left(\frac{2 \rho^2}{w^2\left(z\right)}\right) e^{-\frac{\rho^2}{w^2\left(z\right)}} e^{- \frac{i k \rho^2 z}{2\left(z^2+z_R^2\right)}} e^{-i n \phi} e^{i \left(2p + |n| +1\right) \arctan\frac{z}{z_R}},
\end{align}
with $(\rho,\phi,z)$ being the cylindrical coordinates, $n$ the azimuthal index giving the angular momentum, $p$ the radial index and $C$ a normalisation constant. We use the usual optical definitions of the beam waist $w(z) = \omega_0 \sqrt{1 + (z/z_R)^2}$ and Rayleigh range $z_R = \pi \omega_0^2/\lambda$. For $z=0$, the LG modes can be written as
\begin{align}
\mathcal{L}_{pn}\left(\rho,\phi,0\right) = \frac{\tilde{C}}{w_0} e^{- \frac{\rho^2}{w_0^2}} L^n_p \left(\frac{2\rho^2}{w_0^2}\right) \left(\frac{x - i y}{w_0} \right)^n.
\end{align}

The functions $f_H, f_L$ and $f_T$ are ratios of polynomials $g$ and $h$  in  the complex coordinates $u$ and $v$, which, in turn, are functions of the Cartesian coordinates $(x,y,z)$ as  given in the main text. Restricting  
$f_H, f_L$ and $f_T$  to  $z=0$, we obtain  ratios of polynomials $p$ and $q$  in the variables $x$ and $y$. Expanding in LG modes {\em without} the overall Gaussian factor  $\exp( - \rho^2/ {w_0^2})$  
we obtain an  expansions with  coefficients which  are polynomials in the parameter $\alpha \equiv \omega_0/l$.  

In this way, we arrive at the following {\em exact} identities:
\begin{align*}
\left.f_H\right|_{z=0} &= \left.\frac{2 \alpha \mathcal{L}_{0 -1}}{i\left[\left(\frac{\alpha^2}{2}-1\right)\mathcal{L}_{00} - \frac{\alpha^2}{2} \mathcal{L}_{10}\right] }\right|_{z=0}, \\
\left. f_L\right|_{z=0} & = \left. \frac{(2 \alpha)^2 \mathcal{L}_{0 -2}}{(2 \alpha)^2 \mathcal{L}_{0 -2}  -\left(- \frac{\alpha^4}{2} + \alpha^2 -1\right)\mathcal{L}_{00} - \left(\alpha^4 - \alpha^2\right)\mathcal{L}_{10} + \frac{\alpha^4}{2} \mathcal{L}_{20}}\right|_{z=0},
\\
\left.f_T\right|_{z=0}& =\left. \frac{(2 \alpha)^3 \mathcal{L}_{0 -3}}{(2 \alpha)^3 \mathcal{L}_{0 -3} + \frac{1}{4} \left[\left(-4 +2\alpha^2 + 2\alpha^4 - 3\alpha^6 \right)\mathcal{L}_{00} + \alpha^2 \left(-2 -4\alpha^2  +9\alpha^4\right)\mathcal{L}_{1 0} + \left(2\alpha^4 - 9\alpha^6\right) \mathcal{L}_{2 0} + 3 \alpha^6 \mathcal{L}_{3 0}\right]}\right|_{z=0}.
\end{align*}
Dropping the restriction on the  expressions on the right-hand-side  to $z=0$ defines  the approximations $\zeta_H,\zeta_L$ and $\zeta_T$  to the functions $f_H,f_L$ and $f_T$  which we used in the main text. 
The coefficients $c_i$ and $c_i^\prime$ used there  are defined by the above expansions.

As discussed in the main text, the $\Lambda$-configuration synthetic magnetic fields are  obtained using  two laser beams with Rabi frequencies $\kappa_1$  and $  \kappa_2$  given by the numerator and denominator of $\zeta_H,\zeta_L$ and $\zeta_T$. In all cases, $\zeta= \kappa_1/\kappa_2$  provides a physically realisable  approximation to  the given function $f$ and yields synthetic magnetic field lines whose topology agrees with that of the level curves of  the complex function $f$.

\end{widetext}


%

\end{document}